\documentclass[11pt]{article}

\usepackage[a4paper,margin=1in]{geometry}
\usepackage{amsmath,amssymb,mathtools}
\usepackage{bm}
\usepackage{graphicx}
\usepackage{booktabs}
\usepackage{array}
\usepackage{cite}
\usepackage{xcolor}
\usepackage{url}
\usepackage[font=small]{caption}

\makeatletter
\@ifundefined{IEEEkeywords}{%
  \newenvironment{IEEEkeywords}
    {\par\small\noindent\textbf{Index Terms---}}
    {\par}%
}{}
\@ifundefined{IEEEpubid}{\newcommand{\IEEEpubid}[1]{}}{}
\@ifundefined{appendices}{\newcommand{\appendices}{\appendix}}{}
\makeatother

\newcommand{\jj}{\mathrm{j}}
\newcommand{\dd}{\mathrm{d}}
\newcommand{\cH}{\mathsf{H}}
\newcommand{\diag}{\operatorname{diag}}

\newcommand{\Null}{\operatorname{null}}
\newcommand{\proj}{\operatorname{proj}}

\newcommand{\argmin}{\operatorname*{arg\,min}}

\newcommand{\oslashop}{\mathbin{\oslash}}

\title{Shared-State Inverse Synthesis of Dual-Mode Conformal Metasurfaces for Low Scattering and Telemetry}

\author{%
Mohamed R. Khodja\thanks{Corresponding author. E-mail: mkhodja@pmu.edu.sa.} \hspace{0.1em} and Ghala A. Alomar\\[0.5em]
\small Department of Electrical Engineering, College of Engineering,\\
\small Prince Mohammad Bin Fahd University, Khobar 31952, Saudi Arabia
}

\date{}

\begin{document}
\maketitle

\begin{abstract}
A conformal surface that alternates between sectoral low scattering and directive telemetry poses a shared-hardware inverse problem when both modes must use one finite passive impedance alphabet, although their spatial codebooks may differ. This paper formulates and solves that mixed continuous--discrete problem at the effective impedance-boundary level. Reference-centered Tikhonov preimages prescribe restricted radiation data; inverse-source null-space freedom then moves each current within its data-equivalent class toward impedance profiles quantizable by the same alphabet. A line-searched Gauss--Newton update, common-state assignment and projected-centroid updates, and exact forward refinement determine the currents, two codebooks, and one passive library. Data invariance, the current-to-impedance differential, the optimal state update, monotonic mismatch reduction, and a forward perturbation bound are derived. A full-mutual-coupling two-dimensional TM study of a $4\lambda_0$-diameter cylinder uses 64 physical cells and one shared eight-state library. On a 256-unknown validation model, the worst power ratio over $\pm10^{\circ}$ monostatic sectors and incidence angles from $-30^{\circ}$ to $30^{\circ}$ is $-11.6$~dB relative to a PEC cylinder; telemetry directivity is 13.1~dB with $-13.6$-dB peak sidelobes and a $1^{\circ}$ pointing error. A separate Fourier--Galerkin formulation, converged through 513 complex modes without reoptimization, gives $-11.36$~dB and 13.02~dB for the corresponding worst-sector and directivity metrics. The results establish feasibility at the effective impedance-boundary level; they do not establish unit-cell, bias-network, or platform-level realizability.
\end{abstract}


\section{Introduction}

\IEEEpubid{}
Metasurfaces provide a thin electromagnetic boundary whose local response can be patterned to transform incident fields \cite{Holloway2012Overview}. Passive conformal beamformers can be synthesized from field transformations and full-wave surface models \cite{Budhu2022Beamform}, while current-based procedures can enforce radiation masks and retrieve realizable impedance profiles \cite{Zucchi2023Current,Teodorani2024Generalized}. Reconfigurable impedance surfaces extend the same principle to electronically selected scattering states \cite{Sievenpiper2002Tunable,Cui2014Coding,Wen2023RealData}. Separately, phase-coded surfaces have long been used to redirect rather than absorb monostatic scattering \cite{Paquay2007AMC,Chen2015Checkerboard}.

The two functions are attractive on a common aerospace aperture: a low-scattering state limits energy returned to a specified threat sector, whereas a telemetry state forms a directive communication beam. The electromagnetic objectives are not logically contradictory because the low-scattering requirement is sectoral, not a demand for zero scattering in every direction. For a passive object, energy removed from a protected sector must be dissipated or redistributed. The practical conflict is instead architectural: the same cells, bias network, and finite set of load states must support both current distributions.

Inverse-source nonuniqueness offers a useful degree of freedom. Far-field constraints determine only the radiating component of a surface current; currents in the null space of the restricted radiation operator do not alter the constrained samples. This general observation, and its use in metasurface synthesis, are not claimed as new here.

The closest recent formulation is that of Oliveri \emph{et al.} \cite{Oliveri2025InverseSource}. Their method decomposes the current into preimage and null-space components and alternately optimizes the null-space coefficients and the layout of one inexpensive static passive planar electromagnetic skin for a prescribed reflected pattern. It is linked to an inexpensive meta-atom database and is supported by a printed paper-substrate prototype. The present problem is different in its coupling constraint: one conformal reconfigurable surface must support low-scattering and telemetry modes, the low-scattering code must serve several illuminations, and all mode-dependent codebooks must draw from the same finite passive impedance alphabet. Thus, the methodological overlap is the use of inverse-source freedom, whereas the contribution here is the shared-state, dual-mode compatibility problem and its solution. The two studies are complementary rather than alternatives, and the present work deliberately stops at the effective impedance-boundary level.

The main contributions are as follows.

\begin{enumerate}
\item A mixed continuous--discrete inverse problem is introduced in which different low-scattering and telemetry codebooks, including one low-scattering code shared across multiple illuminations, must use one passive $K$-state impedance alphabet.
\item A solution strategy uses inverse-source null-space freedom to reconcile the data-equivalent currents with that shared alphabet. The Fr\'{e}chet differential from current perturbations to retrieved impedance yields a regularized, line-searched Gauss--Newton step.
\item Closed-form assignment and projected-centroid updates are obtained for the common library. Observation invariance, monotonic mismatch reduction, and a resolvent-based robustness bound are proved.
\item The frozen design is evaluated with a full-mutual-coupling impedance-boundary integral equation and independently cross-validated by a Fourier--Galerkin formulation with a different basis, testing procedure, and representation of the discontinuous impedance profile.
\end{enumerate}

The contribution is theoretical and computational. The local scalar impedance model deliberately separates the shared-state inverse question from unit-cell geometry, polarization conversion, bias routing, and thermo-mechanical construction. Accordingly, the results establish feasibility at the effective impedance-boundary level; they do not establish unit-cell, bias-network, or platform-level realizability.

\section{Electromagnetic and Hardware Model}

\subsection{Surface Equation and Far-Field Operator}

The exterior medium is free space, with wavenumber \(k=2\pi/\lambda_0=\omega\sqrt{\mu_0\epsilon_0}\) and intrinsic impedance \(\eta=\sqrt{\mu_0/\epsilon_0}\). Fields have the $e^{-\jj\omega t}$ time dependence. On a smooth conformal boundary $\Gamma$, a scalar impedance boundary condition is written abstractly as
\begin{equation}
 \left[\mathcal{S}_{\Gamma}+\mathcal{D}(Z_m)\right]J_{m,r}=b_{m,r},
 \label{eq:generalIBC}
\end{equation}
where $m\in\{\mathrm{S},\mathrm{T}\}$ denotes the low-scattering or telemetry mode, $r$ indexes an excitation scenario, $J_{m,r}$ is an equivalent tangential electric current, $b_{m,r}$ is the impressed tangential electric field, and $\mathcal{D}(Z)$ is multiplication by the local sheet impedance. Equation~\eqref{eq:generalIBC} also describes a three-dimensional vector formulation after replacing the scalar quantities by tangential vectors and dyadics. The synthesis below depends only on the discretized operators and therefore carries to that setting.

For a reproducible canonical study, $\Gamma$ is specialized to a two-dimensional circular cylinder and TM$_z$ polarization. The local impedance condition follows the standard approximate-boundary framework \cite{Senior1995ABC}. The exterior field radiated by $J_z$ is
\begin{equation}
 E_z^{\mathrm{sca}}(\bm r)=-\frac{k\eta}{4}
 \int_{\Gamma}H_0^{(1)}\!\left(k\lvert\bm r-\bm r'\rvert\right)
 J_z(\bm r')\,\dd s',
 \label{eq:singlelayer}
\end{equation}
where \(\bm r,\bm r'\) are the observation and source points, respectively, \(H_0^{(1)}\) is Hankel function of the first kind and order zero, and \(J_z\) is the \(z\)-directed equivalent electric surface-current density on \(\Gamma\). Hence $E_z^{\mathrm{inc}}+E_z^{\mathrm{sca}}=ZJ_z$ becomes
\begin{equation}
 \left[\bm D(\bm z)+\bm S\right]\bm j=\bm b,
  \label{eq:discreteForward}
\end{equation}
where \(\bm z,\bm j,\bm b\) are the discrete impedance, current, and excitation vectors obtained after MoM discretization, \(\bm S\) is the dense MoM interaction matrix representing the discretized single-layer operator, and \(\bm D(\bm z)=\diag(\bm z)\), i.e., the diagonal matrix whose entries are the components of \(\bm z\). Pulse bases and midpoint testing are used on equal arc-length panels of length $\Delta s$ \cite{Harrington1993Moment}. For $p\ne q$,
\begin{equation}
 S_{pq}=\frac{k\eta\Delta s}{4}H_0^{(1)}(kR_{pq}),
 \label{eq:offdiag}
\end{equation}
where \(R_{pq}\) is the distance between the testing point of panel \(p\) and the source point or midpoint of panel \(q\). The logarithmic self term follows from
\begin{align}
 \int_{-\Delta s/2}^{\Delta s/2}H_0^{(1)}(k\lvert u\rvert)\,\dd u
 &=\Delta s\left[1+\frac{2\jj}{\pi}
 \left(\ln\frac{k\Delta s}{4}+\gamma-1\right)\right]
 \nonumber\\[-1mm]
 &\quad+\mathcal{O}\!\left((k\Delta s)^2\ln(k\Delta s)\right),
 \label{eq:selfterm}
\end{align}
where \(\gamma\) is the Euler–Mascheroni constant. The far-field angular factor sampled at $\theta_\ell$ is
\begin{equation}
 f_\ell=-\frac{k\eta}{4}\sum_{n=1}^{N}j_n\Delta s
 e^{-\jj k\widehat{\bm r}_{\ell}\cdot\bm r_n}+f_\ell^{\mathrm{dir}}
 = [\bm C\bm j]_\ell+f_\ell^{\mathrm{dir}},
 \label{eq:farfield}
\end{equation}
where \(N\) is the number of MoM current unknowns, \(\widehat{\bm r}_{\ell}\) is the unit observation vector associated with angle \(\theta_\ell\), \(\bm r_n\) is the position of the midpoint of panel \(n\), \(\bm C\) is the discrete radiation operator mapping the current vector to the selected far-field samples, and \(f_\ell^{\mathrm{dir}}\) is the direct-source contribution at observation angle \(\theta_\ell\). The direct term is zero for plane-wave scattering and is retained for the telemetry line source. Equations~\eqref{eq:singlelayer}--\eqref{eq:farfield} include mutual coupling; they are not a local phase-only array factor. The impedance is normalized by $\eta$, so all tabulated values are dimensionless.

\subsection{One Library, Two Codebooks}

Let
\begin{equation}
 \mathcal{Z}(\bm\zeta)=\{\zeta_1,\ldots,\zeta_K\},\qquad
 \bm\zeta\in\mathcal{P}^K
 \label{eq:library}
\end{equation}
be the states physically available in every cell. Passivity is imposed through the closed convex set
\begin{equation}
 \mathcal{P}=\{R_0+\jj X:X_{\min}\leq X\leq X_{\max},\ R_0\geq0\}.
 \label{eq:passiveSet}
\end{equation}
The modes use distinct integer codebooks
\begin{equation}
 q_m(n)\in\{1,\ldots,K\},\qquad
 z_{m,n}=\zeta_{q_m(n)},\qquad m\in\{\mathrm S,\mathrm T\},
 \label{eq:codebooks}
\end{equation}
but $\bm\zeta$ is common. This distinction is essential: switching all cells is allowed, manufacturing two mode-specific sets of meta-atoms is not.

For plane-wave incidence $\alpha_r$, define the protected observation sector
\begin{equation}
 \Omega_r=\{\theta:\lvert\operatorname{wrap}(\theta-\alpha_r-\pi)\rvert
 \leq\Delta_\mathrm S\},
\end{equation}
where \(\operatorname{wrap}(\cdot)\) stands for the angular wrapping into a stated interval and \(\Delta_{\mathrm S}\) is the half-width of the protected low-scattering sector. The sectoral power ratio relative to a PEC cylinder is
\begin{align}
 \rho_r&=
 \frac{\max_{\theta\in\Omega_r}
 \lvert f_{\mathrm S,r}(\theta)\rvert^2}
 {\max_{\theta\in\Omega_r}
 \lvert f_{\mathrm{PEC},r}(\theta)\rvert^2},
 \nonumber\\
 \bar{\rho}&=\frac{1}{R}\sum_{r=1}^{R}\rho_r,
 \qquad
 \rho_{\max}=\max_{1\leq r\leq R}\rho_r,
 \nonumber\\
 \Phi_{\mathrm S}&=0.45\bar{\rho}+0.55\rho_{\max}.
 \label{eq:stealthObjective}
\end{align}
For reporting, the corresponding decibel quantities are
$\rho_{r,\mathrm{dB}}=10\log_{10}\rho_r$,
$\bar{\rho}_{\mathrm{dB}}=10\log_{10}\bar{\rho}$, and
$\rho_{\max,\mathrm{dB}}=10\log_{10}\rho_{\max}$.
Let \(P(\theta)=\lvert f_{\mathrm T}(\theta)\rvert^2\) be the continuous power pattern. For telemetry, let $\mathcal M$ be the $\pm8^{\circ}$ main-beam sector and $\mathcal L$ the region outside $\pm15^{\circ}$. With $P_\ell=P(\theta_\ell)=\lvert f_{\mathrm T}(\theta_\ell)\rvert^2$, let $v_\ell=1$ except for the two samples at the boundaries of $\mathcal M$, where $v_\ell=1/2$.
\begin{align}
 \eta_{\mathcal M}&=\frac{\sum_{\ell\in\mathcal M}v_\ell P_\ell}{\sum_\ell P_\ell},\nonumber\\
 \rho_{\mathcal L}&=\frac{\max_{\ell\in\mathcal L}P_\ell}{\max_{\ell\in\mathcal M}P_\ell},
 \label{eq:teleMetrics}\\
 \Phi_{\mathrm T}&=1-\eta_{\mathcal M}+0.15\rho_{\mathcal L}
 +0.002\left(\frac{\theta_{\max}-\theta_0}{1^{\circ}}\right)^2.
 \label{eq:teleObjective}
\end{align}
Here, \(\theta_{\max}=\arg\max_{\theta}P(\theta)\) is the direction of the telemetry-pattern maximum, and \(\theta_0\) is the prescribed telemetry-beam direction. The sectoral normalization follows standard RCS comparison practice \cite{Knott2004RCS}; the telemetry directivity convention is two-dimensional \cite{Balanis2016Antenna}. The two-dimensional directivity is
\[
D_{\mathrm{2D}}
=10\log_{10}\!\left[
\frac{2\pi\max_{\theta}P(\theta)}
{\displaystyle\int_{0}^{2\pi}P(\theta)\,\mathrm d\theta}
\right],
\]
and the peak sidelobe level is \(\mathrm{SLL}=10\log_{10}\rho_{\mathcal L}\). Both quantities are reported in decibels. The final forward polish minimizes $4\Phi_{\mathrm S}+0.7\Phi_{\mathrm T}$. The coefficients were fixed before the reported optimization and scale the two dimensionless objectives to comparable initial magnitudes.

\section{Shared-State Inverse Synthesis}

\subsection{Reference-Centered Preimage}

For one scenario, collect the constrained current-generated far-field samples in $\bm C\bm j$ and subtract any direct field from the desired constrained far-field-data vector $\bm d$ after accounting for any direct field. Starting from a physically induced reference current $\bm j_{\mathrm{ref}}$, consider
\begin{equation}
 \min_{\bm j}\ \lVert\bm C\bm j-\bm d\rVert_2^2
 +\lambda\lVert\bm j-\bm j_{\mathrm{ref}}\rVert_2^2,\qquad \lambda>0.
 \label{eq:tikhonov}
\end{equation}
Its unique solution is
\begin{equation}
 \bm j_{\mathrm p}=\bm j_{\mathrm{ref}}+
 \bm C^{\cH}(\bm C\bm C^{\cH}+\lambda\bm I)^{-1}
 (\bm d-\bm C\bm j_{\mathrm{ref}}), 
 \label{eq:preimage}
\end{equation}
where \((\cdot)^{\mathsf H}\) denotes the conjugate transpose and \(\bm I\) is the identity matrix of the required dimension. The singular values of $\bm C$ expose the conditioning of the restricted inverse source problem, and $\lambda$ suppresses corrections in poorly observed directions \cite{Colton2019Inverse,Hansen1992LCurve,Hansen1998Rank,Tikhonov1977IllPosed,Golub2013Matrix}.

\textit{Proposition 1 (data invariance):} Let the columns of $\bm N$ span $\Null(\bm C)$. Then every
\begin{equation}
 \bm j(\bm a)=\bm j_{\mathrm p}+\bm N\bm a
 \label{eq:nullfamily}
\end{equation}
has the same constrained samples as $\bm j_{\mathrm p}$. Let \(\tau\) be the upper bound on the approximate null-space residual \(\lVert\bm C\bm N\rVert_2\), i.e., $\lVert\bm C\bm N\rVert_2\leq\tau$, then
\begin{equation}
 \lVert\bm C\bm j(\bm a)-\bm C\bm j_{\mathrm p}\rVert_2
 \leq\tau\lVert\bm a\rVert_2.
 \label{eq:approxNull}
\end{equation}

\textit{Proof:} Substitution into $\bm C\bm j(\bm a)$ gives $\bm C\bm j_{\mathrm p}+\bm C\bm N\bm a$. The exact result follows from $\bm C\bm N=\bm0$; the approximate result follows from submultiplicativity. Equation~\eqref{eq:preimage} follows independently from the normal equation of \eqref{eq:tikhonov} and the push-through identity $(\bm C^{\cH}\bm C+\lambda\bm I)^{-1}\bm C^{\cH}=\bm C^{\cH}(\bm C\bm C^{\cH}+\lambda\bm I)^{-1}$. \hfill$\square$

\subsection{Current-to-Impedance Differential}

For a current with nonzero entries, the impedance that makes it satisfy \eqref{eq:discreteForward} is
\begin{equation}
 \bm z(\bm j)=(\bm b-\bm S\bm j)\oslashop\bm j,
 \label{eq:retrieve}
\end{equation}
where the operator \(\oslashop\) denotes elementwise division. Entries below a prescribed current floor are excluded or regularized. This is necessary: impedance retrieval is intrinsically unstable at current zeros.

\textit{Proposition 2 (retrieval differential):} The directional differential of \eqref{eq:retrieve} is
\begin{equation}
 \mathrm D\bm z[\bm j](\delta\bm j)=
 -\bm D(\bm 1 \oslashop\bm j)
 \left[\bm S+\bm D(\bm z(\bm j))\right]\delta\bm j.
 \label{eq:jacobian}
\end{equation}

\textit{Proof:} For component $n$, the quotient rule gives
\begin{align}
 \delta z_n&=-\frac{[\bm S\delta\bm j]_n}{j_n}
 -\frac{b_n-[\bm S\bm j]_n}{j_n^2}\delta j_n \nonumber\\
 &=-\frac{1}{j_n}\left([\bm S\delta\bm j]_n+z_n\delta j_n\right),
\end{align}
which stacked over $n$ yields \eqref{eq:jacobian}. \hfill$\square$

With $\delta\bm j=\bm N\bm a$, define
\begin{equation}
 \bm B(\bm j)=-\bm D(1\oslashop\bm j)
 [\bm S+\bm D(\bm z)]\bm N.
 \end{equation}
For fixed states and assignments, a null-space update is obtained from
\begin{equation}
 \min_{\bm a}\ \left\lVert\bm W
 [\bm B\bm a-(\bm\zeta_{\bm q}-\bm z)]\right\rVert_2^2
 +\mu\lVert\bm a\rVert_2^2.
 \label{eq:nullGN}
\end{equation}
Here, \(\bm W\) is the diagonal weighting matrix used in the impedance-mismatch problem. Specifically, \(\mathbf W=\operatorname{diag}(\sqrt{w_1},\ldots,\sqrt{w_N})\), where \(w_n\geq0\) is the weight assigned to the \(n\)th retrieved impedance sample. For low-scattering scenario \(r\), \(w_n=w_{r,n}\), whereas for the telemetry mode, \(w_n=w_{\mathrm T,n}\). Also, $\bm\zeta_{\bm q}=[\zeta_{q(1)},\ldots,\zeta_{q(N)}]^T$, and \(\mu\) is the nonnegative Tikhonov or step-regularization parameter for the null-space update. The solution is
\begin{equation}
 \bm a=(\bm B^{\cH}\bm W^2\bm B+\mu\bm I)^{-1}
 \bm B^{\cH}\bm W^2(\bm\zeta_{\bm q}-\bm z).
 \label{eq:nullClosed}
\end{equation}
A backtracking line search on $\bm j+t\bm N\bm a$ enforces the current floor and accepts only a decrease of the exact, not linearized, impedance mismatch. Parameter \(t\) is the backtracking step length.

\subsection{Common-State Quantization}

Let $\bm z_{\mathrm S,r}$ be the profiles retrieved for the $R$ low-scattering illuminations and $\bm z_{\mathrm T}$ the telemetry profile. One low-scattering assignment is shared by all $R$ illuminations. For fixed states,
\begin{align}
 q_{\mathrm S}(n)&=\argmin_k\sum_{r=1}^{R}w_{r,n}
 \lvert z_{\mathrm S,r,n}-\zeta_k\rvert^2,\label{eq:assignS}\\
 q_{\mathrm T}(n)&=\argmin_k w_{\mathrm T,n}
 \lvert z_{\mathrm T,n}-\zeta_k\rvert^2.
 \label{eq:assignT}
\end{align}
These assignments couple the scenarios through the code, while the library couples both modes.

\textit{Proposition 3 (optimal state update):} For fixed assignments, let $\mathcal I_k$ contain all weighted impedance samples assigned to state $k$. The minimizer over $\mathcal P$ is
\begin{equation}
 \zeta_k^{+}=\proj_{\mathcal P}\left(
 \frac{\sum_{i\in\mathcal I_k}w_i z_i}{\sum_{i\in\mathcal I_k}w_i}\right).
 \label{eq:centroid}
\end{equation}

\textit{Proof:} Completing the square separates the objective for state $k$ into a positive weight times $\lvert\zeta_k-\bar z_k\rvert^2$ plus a constant. Minimization on a closed convex set is therefore the Euclidean projection of the weighted centroid $\bar z_k$. \hfill$\square$

This is a constrained complex version of least-squares quantization \cite{Lloyd1982Quantization}. For \eqref{eq:passiveSet}, projection fixes $\Re\{\zeta_k\}=R_0$ and clips $\Im\{\zeta_k\}$ to $[X_{\min},X_{\max}]$.

\textit{Proposition 4 (monotonic inverse stage):} Suppose every null-space step is accepted only when it decreases the exact mismatch, and empty clusters retain their previous states. Then the sequence of shared-library mismatch values produced by \eqref{eq:nullGN}, \eqref{eq:assignS}--\eqref{eq:assignT}, and \eqref{eq:centroid} is nonincreasing and convergent.

\textit{Proof:} The assignment update selects the minimum among $K$ choices at each cell. The state update is the global minimizer of each fixed-cluster quadratic by Proposition~3. The line search rejects every current update that increases the exact mismatch. Each block therefore cannot increase the objective. The objective is nonnegative, so its scalar sequence converges. This statement does not assert convergence to a global minimizer or convergence of the iterates. \hfill$\square$

\subsection{Forward Verification and Efficient Discrete Refinement}

The inverse stage minimizes realizability error, not the final electromagnetic objectives. Every candidate is therefore re-solved through \eqref{eq:discreteForward}. Let \(\bm e_n\) be the \(n\)th canonical coordinate vector. Changing cell $n$ from $z_n$ to $z_n+\delta z$ gives
\begin{equation}
 \bm M'=\bm M+\delta z\bm e_n\bm e_n^T,
 \qquad \bm M=\bm D(\bm z)+\bm S.
\end{equation}
Provided $1+\delta z[\bm M^{-1}]_{nn}\ne0$, the inverse update is
\begin{equation}
 \bm M'^{-1}=\bm M^{-1}-
 \frac{\delta z\bm M^{-1}\bm e_n\bm e_n^T\bm M^{-1}}
 {1+\delta z\bm e_n^T\bm M^{-1}\bm e_n},
 \label{eq:SM}
\end{equation}
which is the Sherman--Morrison formula \cite{Sherman1950Inverse}. It makes exhaustive $K$-state coordinate trials inexpensive. The shared state reactances are subsequently polished by a bounded derivative-free search, followed by a final coordinate sweep \cite{Bertsekas2016Nonlinear}. Each forward update is retained only if it reduces \eqref{eq:stealthObjective} or \eqref{eq:teleObjective}; consequently, the reported codebooks are exactly feasible for the shared library.

The complete synthesis is:
\begin{enumerate}
\item choose observation masks, reference currents, $K$, $\lambda$, and the passive set $\mathcal P$;
\item compute \eqref{eq:preimage}, construct the null projector of each restricted radiation operator, and extract a numerical basis of $\Null(\bm C)$;
\item retrieve impedances using \eqref{eq:retrieve};
\item alternate \eqref{eq:assignS}--\eqref{eq:centroid} with the line-searched update \eqref{eq:nullClosed};
\item solve the physical forward problems \eqref{eq:discreteForward};
\item polish the common library and both codebooks using exact forward objectives and \eqref{eq:SM}; and
\item freeze the design and evaluate it on independent finer meshes and perturbed parameters.
\end{enumerate}

\section{Stability of the Forward Realization}

Quantization error, fabrication error, and model discrepancy perturb $\bm M$. The following bound gives a direct stability diagnostic.

\textit{Proposition 5 (current and far-field perturbation):} Let $\widetilde{\bm M}=\bm M+\Delta\bm M$, $\bm j=\bm M^{-1}\bm b$, and $\widetilde{\bm j}=\widetilde{\bm M}^{-1}\bm b$. If
\begin{equation}
 \epsilon=\lVert\bm M^{-1}\Delta\bm M\rVert_2<1,
 \end{equation}
then
\begin{align}
 \frac{\lVert\widetilde{\bm j}-\bm j\rVert_2}{\lVert\bm j\rVert_2}
 &\leq\frac{\epsilon}{1-\epsilon},\label{eq:currentBound}\\
 \lVert\bm C(\widetilde{\bm j}-\bm j)\rVert_2
 &\leq\lVert\bm C\rVert_2\frac{\epsilon}{1-\epsilon}\lVert\bm j\rVert_2.
 \label{eq:fieldBound}
\end{align}

\textit{Proof:} Since $\widetilde{\bm M}=\bm M(\bm I+\bm M^{-1}\Delta\bm M)$,
\begin{equation}
 \widetilde{\bm j}-\bm j=-
 (\bm I+\bm M^{-1}\Delta\bm M)^{-1}
 \bm M^{-1}\Delta\bm M\bm j.
\end{equation}
The Neumann-series bound gives $\lVert(\bm I+\bm K)^{-1}\rVert_2\leq(1-\lVert\bm K\rVert_2)^{-1}$ for $\lVert\bm K\rVert_2<1$, proving \eqref{eq:currentBound}; multiplication by $\bm C$ proves \eqref{eq:fieldBound}. \hfill$\square$

The sufficient condition is conservative, particularly near a resonance. It nevertheless explains why impedance mismatch alone is not enough: a small mismatch can produce a large field change when $\bm M$ is poorly conditioned. Exact forward verification and random perturbation tests are therefore part of the synthesis, not optional post-processing.

\section{Numerical Study}

\subsection{Configuration and Verification Protocol}

Table~\ref{tab:configuration} lists the adopted parameters. Length is normalized by \(\lambda_0\), and all impedances are normalized by the free-space impedance \(\eta_0\); hence, \(\eta=1\) in the numerical implementation. The circumference contains 64 independently controlled physical cells, each of arc length $0.196\lambda_0$. Two pulse unknowns per physical cell are used during final synthesis. The frozen design is then evaluated with four and six unknowns per cell. A line source at $(1.45\lambda_0,0)$, inside the conformal boundary, excites the telemetry mode toward $\theta_0=0^{\circ}$. Thirteen incidence angles are enforced in the low-scattering mode. The selected eight states are
\begin{equation}
 \frac{\bm\zeta}{\eta}=0.03+\jj\left\{
 \begin{array}{rrrr}
 -6.756,&-1.618,&-0.562,&-0.224,\\
 \phantom{-}0.188,&\phantom{-}0.607,&\phantom{-}1.345,&\phantom{-}4.932
 \end{array}\right\}.
 \label{eq:finalStates}
\end{equation}
All states are passive effective boundary impedances. Their use establishes feasibility within the prescribed impedance-boundary model, not the realizability of a particular lumped or printed cell, its bias network, or its installation on an aerospace platform.

\begin{table}[t]
\caption{Numerical Configuration}
\label{tab:configuration}
\centering
\footnotesize
\begin{tabular}{@{}p{0.42\columnwidth}p{0.52\columnwidth}@{}}
\toprule
Quantity & Adopted value \\
\midrule
Cylinder radius & $a=2\lambda_0$ \\
Physical control cells & 64 ($0.196\lambda_0$ arc length) \\
Synthesis / validation unknowns & 128 / 256 and 384 \\
Shared states & $K=8$, $R_0=0.03\eta$ \\
Low-scattering incidences & $-30^{\circ}:5^{\circ}:30^{\circ}$ \\
Protected sector & $\pm10^{\circ}$ about monostatic direction \\
Telemetry feed & $(a-0.55\lambda_0,0)$ \\
Telemetry target & $\theta_0=0^{\circ}$ \\
Main / sidelobe regions & $\pm8^{\circ}$ / outside $\pm15^{\circ}$ \\
Frequency model & single-frequency, frozen impedance in sweep \\
\bottomrule
\end{tabular}
\end{table}

\begin{figure*}[!tbp]
\centering
\includegraphics[width=0.95\textwidth]{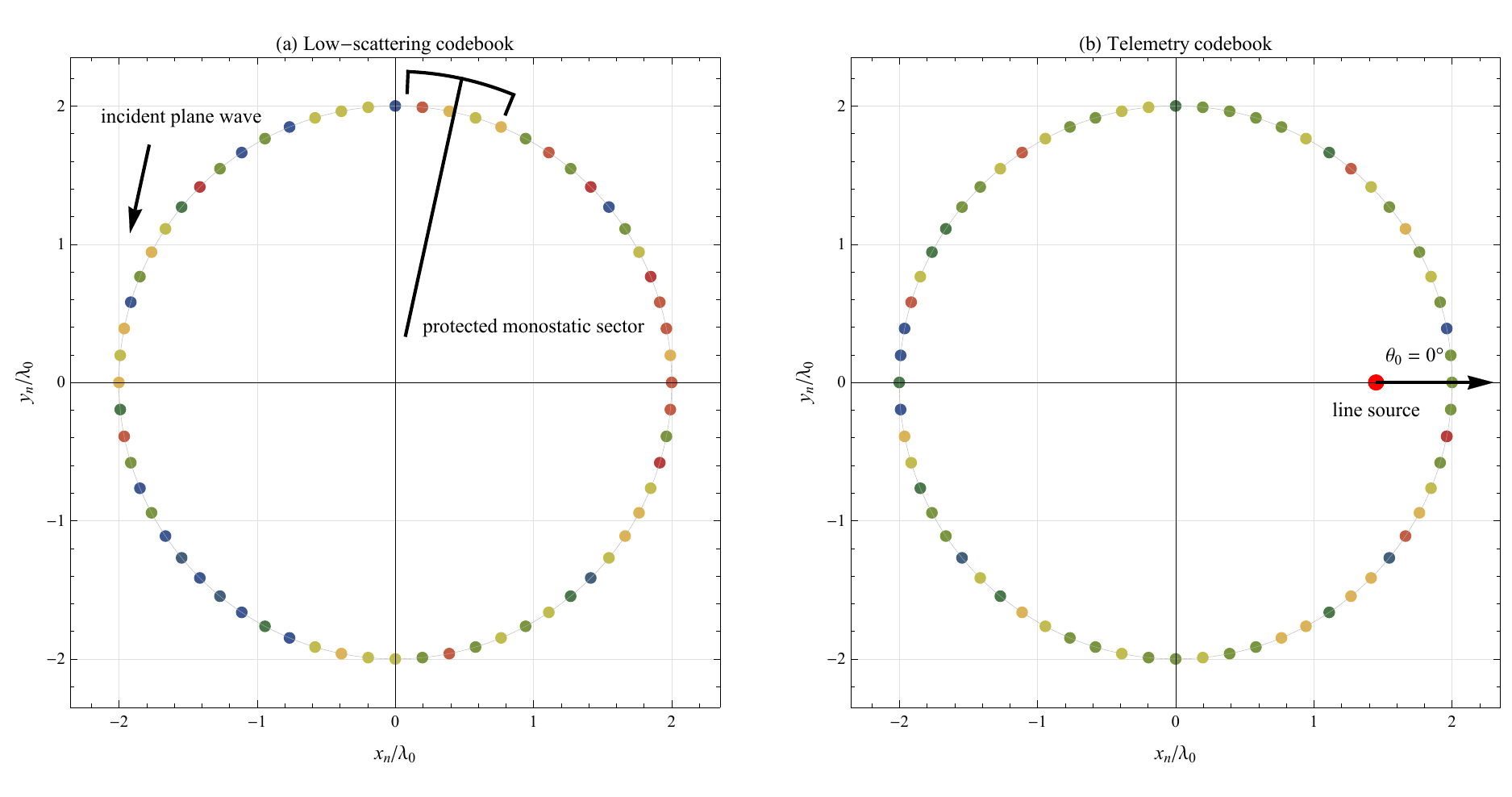}
\caption{Metasurface configuration and mode-dependent codebooks for the transverse cross-section of the two-dimensional $4\lambda_0$-diameter circular cylinder. The boundary is partitioned into 64 independently controlled cells, and color identifies the selected member of the common eight-state impedance alphabet in \eqref{eq:finalStates}. (a) Low-scattering codebook shared by the 13 plane-wave illuminations; the arrows schematically denote a representative incident direction and its associated $\pm10^{\circ}$ protected monostatic sector. (b) Telemetry codebook with the internal line source at $(x_f,y_f)=(1.45\lambda_0,0)$ and the prescribed beam direction $\theta_0=0^{\circ}$. The two modes use different spatial assignments while retaining the same effective impedance-state alphabet.}
\label{fig:configuration}
\end{figure*}

The codebooks are shown in Fig.~\ref{fig:configuration}. Their aperiodic appearance is expected because mutual coupling, curvature, multiple incidence angles, and the shared library are all retained. The numerical implementation first checks the PEC-cylinder solution against the analytic cylindrical-wave series. At 256 unknowns and observation radius $10\lambda_0$, the relative $L_2$ field error is $7.9\times10^{-3}$. The Sherman--Morrison, null-space-invariance, and projector-idempotence errors are $2.6\times10^{-15}$, $2.5\times10^{-16}$, and $7.0\times10^{-16}$, respectively. The low-scattering and telemetry null-space dimensions are 47 and 15, as expected from the sampled operator dimensions, and a one-step inverse smoke test verifies an actual accepted decrease before the production study begins.
The reported main-sector efficiency is an angular power integral evaluated on the uniform observation grid; trapezoidal half weights are used at the two $\pm8^{\circ}$ sector boundaries.

\subsection{Inverse-Source Diagnostics}

\begin{figure*}[!tbp]
\centering
\includegraphics[width=1.00\textwidth]{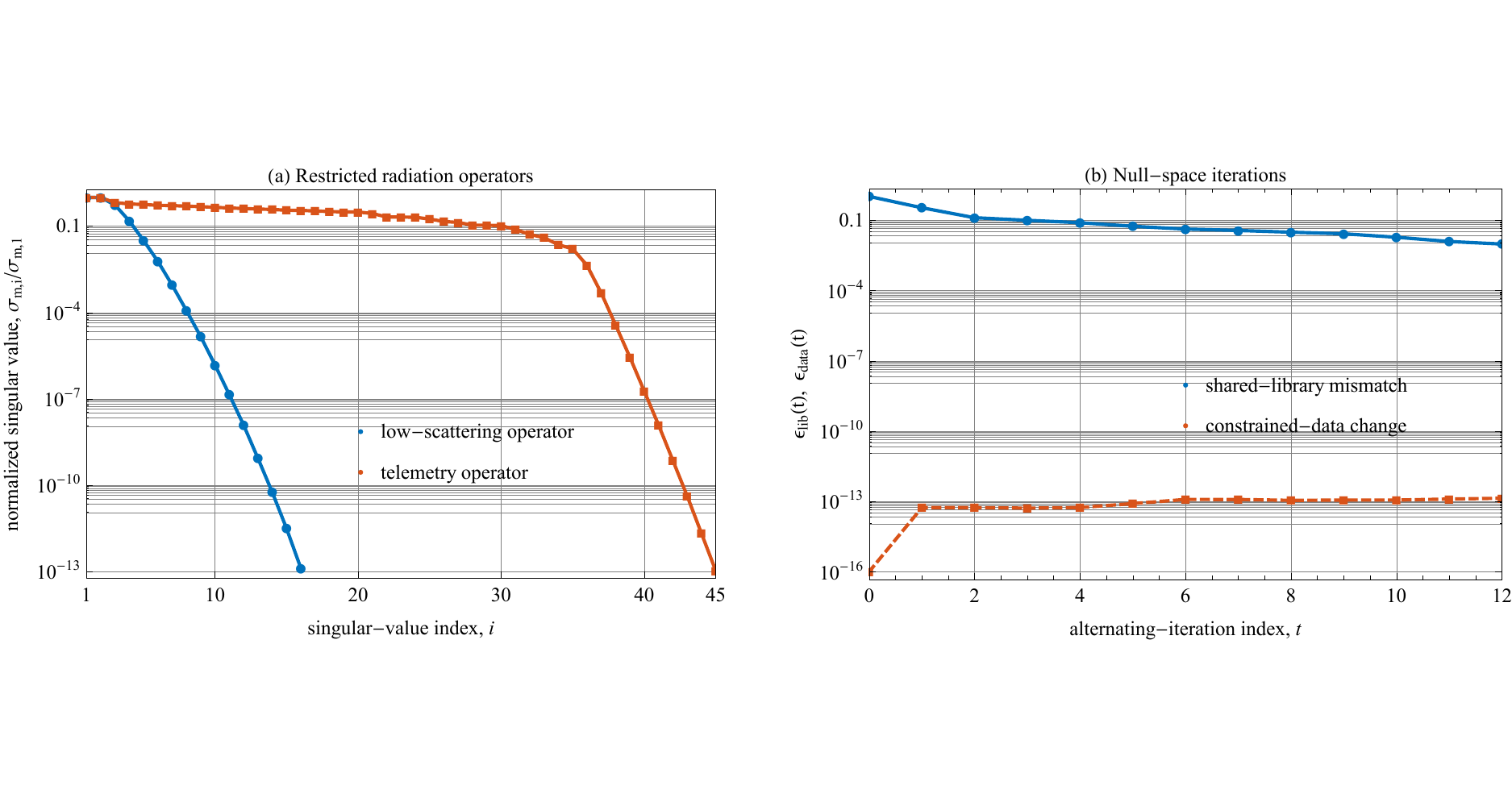}
\caption{Inverse-source diagnostics on the 64-unknown model. (a) Normalized singular-value spectra of the restricted low-scattering and telemetry radiation operators, characterizing the conditioning of the constrained inverse problems and the null-space freedom used by the synthesis. (b) Normalized shared-library mismatch and relative change in the constrained electromagnetic data over 12 alternating null-space/library iterations. The mismatch decreases substantially while the constrained data remain invariant to numerical precision, demonstrating that the retrieved impedances can be brought into agreement with a common finite impedance-state library without materially changing the prescribed observations. This diagnostic uses four initial states; the final forward design uses the eight states in \eqref{eq:finalStates}.}
\label{fig:inverse}
\end{figure*}

Figure~\ref{fig:inverse} isolates the inverse stage on the 64-unknown model. The low-scattering operator samples a $24^{\circ}$ observation sector, and the telemetry operator samples its main beam and distributed sidelobe angles. The rapidly decaying spectra confirm that direct inversion would amplify weakly observed components. Twelve null-space/library iterations reduce the normalized shared-library mismatch from 1.05846 to 0.00978, a 99.1\% decrease. The maximum relative change in the restricted data is $1.37\times10^{-13}$, consistent with Proposition~1 at floating-point precision. The full-row-rank null projector is constructed at high precision, so small but nonzero singular directions remain constrained. This diagnostic uses four initial states to make the realizability mechanism deliberately severe; the final forward design uses the eight shared states in \eqref{eq:finalStates}.

\subsection{Low-Scattering Mode}

\begin{figure*}[!tbp]
\centering
\includegraphics[width=1.00\textwidth]{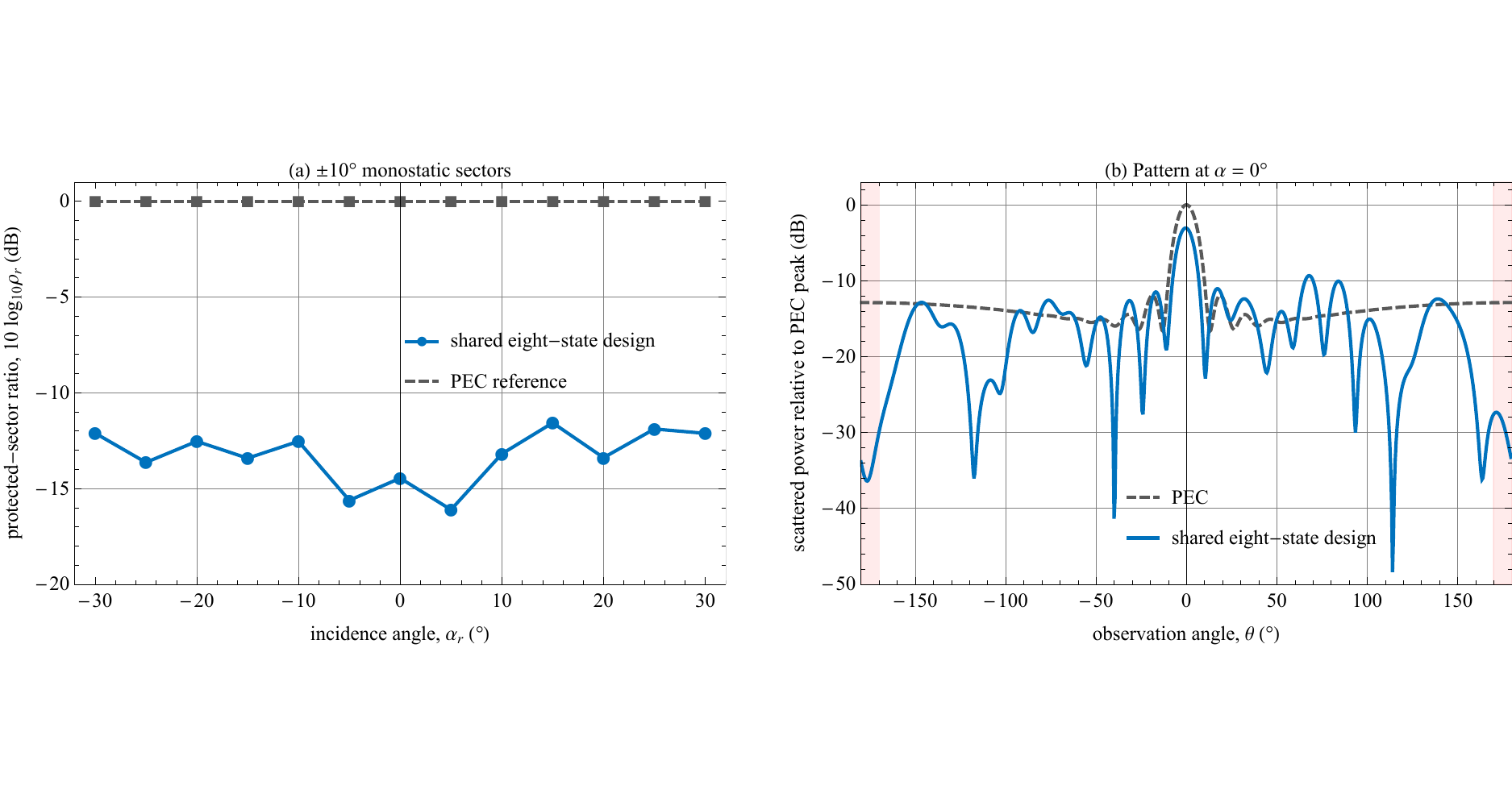}
\caption{Low-scattering performance of the frozen shared eight-state design on the 256-unknown MoM validation mesh. (a) Maximum scattered-power ratio within the $\pm10^{\circ}$ protected monostatic sector, relative to the corresponding PEC-cylinder value, for each of the 13 incidence angles. (b) Full angular scattering pattern for normal incidence, normalized by the PEC peak; the shaded regions mark the wrapped protected sector. The comparison shows that the design reduces the return within the prescribed monostatic sector rather than suppressing scattering uniformly over all observation angles.}
\label{fig:scattering}
\end{figure*}

Figure~\ref{fig:scattering}(a) reports the worst power ratio in each $\pm10^{\circ}$ protected sector on the 256-unknown validation mesh. All 13 incidence samples remain below $-11.5$~dB. The worst and mean ratios are $-11.58$ and $-13.09$~dB. At normal incidence, Fig.~\ref{fig:scattering}(b) shows that the surface redistributes scattering rather than suppressing it everywhere, as required by passivity. Deep nulls outside the protected region are incidental and are not optimization constraints.

\subsection{Telemetry Mode and Cost of Sharing States}

\begin{figure*}[!tbp]
\centering
\includegraphics[width=1\textwidth]{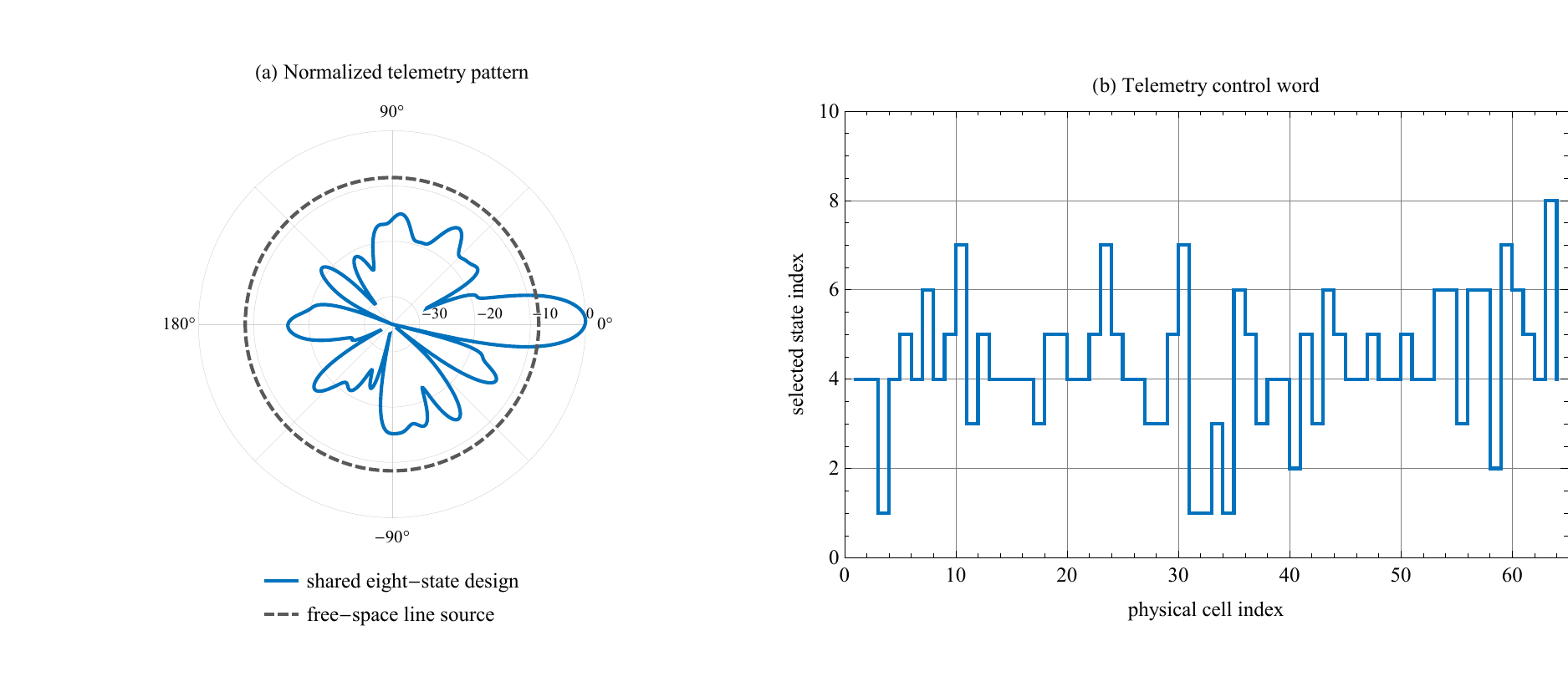}
\caption{Telemetry performance of the frozen shared eight-state design on the 256-unknown MoM validation model. (a) Normalized two-dimensional radiation pattern, including the direct field of the internal line source, with the isotropic free-space line-source response shown for reference. Interaction with the impedance boundary forms a directive beam near the prescribed direction $\theta_0=0^{\circ}$. (b) Telemetry control word, giving the selected shared-state index for each of the 64 physical cells; this integer sequence is the spatial codebook applied in telemetry mode.}
\label{fig:telemetry}
\end{figure*}

The telemetry pattern in Fig.~\ref{fig:telemetry} has 13.10-dB two-dimensional directivity, 0.649 main-sector efficiency, $-13.60$-dB peak sidelobes, and a $1^{\circ}$ pointing error on the 256-unknown model. The free-space two-dimensional line source is isotropic; the directive field is produced by its interaction with the impedance boundary and includes the direct term in \eqref{eq:farfield}.

Table~\ref{tab:performance} distinguishes synthesis-mesh performance from independent validation. The 384-unknown result is close to the 256-unknown result, although the 128-unknown synthesis mesh is optimistic by 2.9~dB in the worst low-scattering metric. Reporting only the synthesis mesh would therefore be misleading. As an upper-bound comparison, two separate eight-state libraries were optimized while retaining the same number of states per mode. At 256 unknowns, separate libraries improve the worst low-scattering ratio by only 0.26~dB and telemetry directivity by 0.15~dB. The shared-library penalty is small, although the mean scattering and sidelobe metrics show a more visible 1.1-dB cost.

\begin{table*}[t]
\caption{Forward-Verified Performance at $f_0$}
\label{tab:performance}
\centering
\footnotesize
\resizebox{\linewidth}{!}{%
\begin{tabular}{@{}lcccccc@{}}
\toprule
Model and library & Worst sector (dB) & Mean sector (dB) & $\eta_{\mathcal M}$ & $D_{\mathrm{2D}}$ (dB) & SLL (dB) & Pointing (deg)\\
\midrule
Shared, 128 unknowns (synthesis) & $-14.29$ & $-15.15$ & 0.653 & 13.07 & $-13.89$ & 0.0\\
Shared, 256 unknowns (validation) & $-11.58$ & $-13.09$ & 0.649 & 13.10 & $-13.60$ & 1.0\\
Shared, 384 unknowns (validation) & $-11.38$ & $-12.88$ & 0.644 & 13.08 & $-13.47$ & 1.5\\
Separate $8+8$ libraries, 256 unknowns & $-11.83$ & $-14.22$ & 0.675 & 13.25 & $-14.74$ & 1.0\\
\bottomrule
\end{tabular}
}
\end{table*}

\subsection{Mesh, Frequency, and Impedance-Error Sensitivity}

\begin{figure*}[!tbp]
\centering
\includegraphics[width=1.00\textwidth]{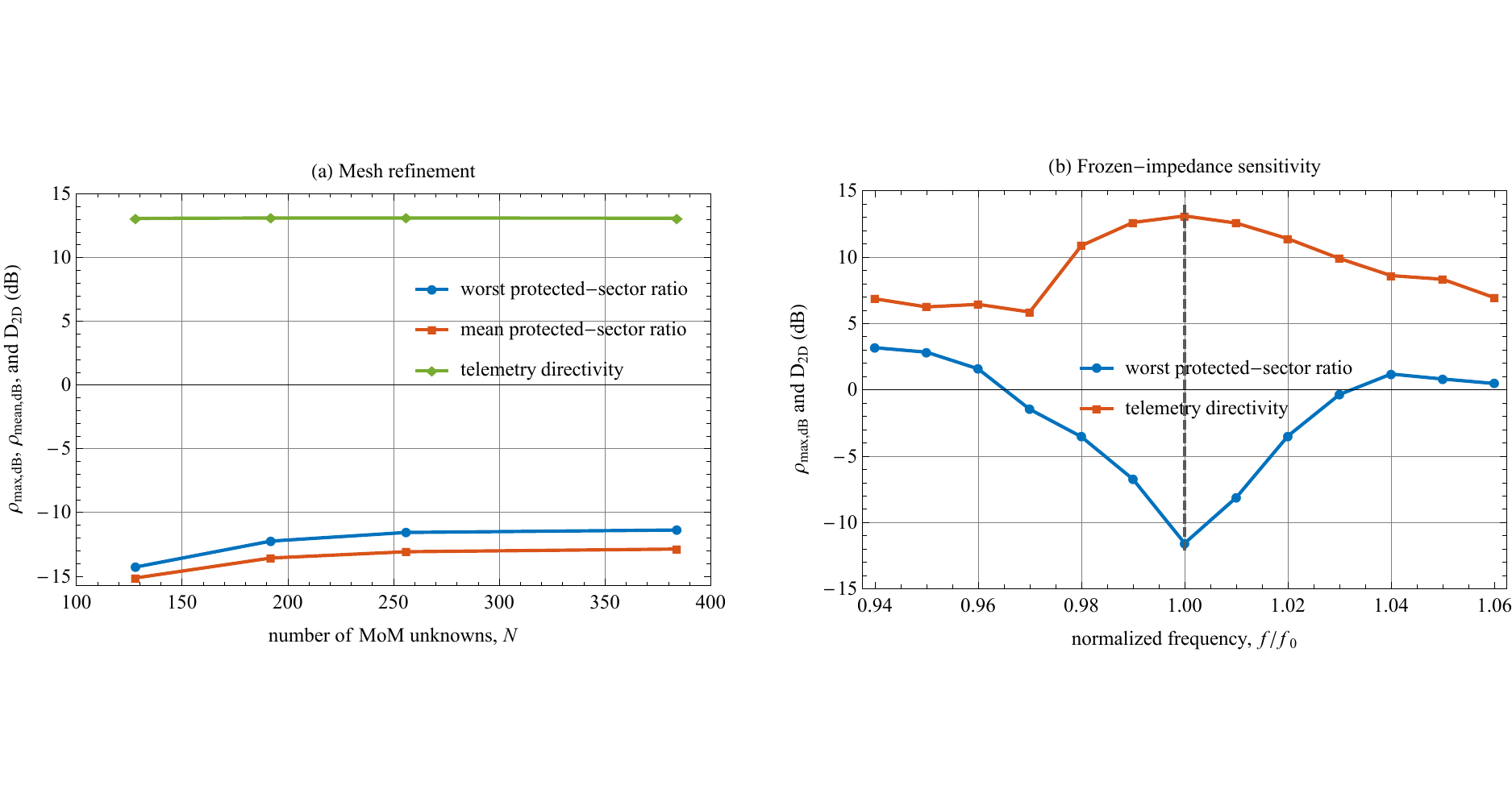}
\caption{Numerical convergence and frequency sensitivity of the frozen 64-cell shared-state design. (a) Worst and mean protected-sector ratios and telemetry two-dimensional directivity versus the number of MoM unknowns, with the physical cells, impedance states, and codebooks held fixed. Convergence toward the 384-unknown result tests sensitivity to the pulse-basis discretization. (b) Worst protected-sector ratio and telemetry directivity obtained by varying frequency while retaining the impedances prescribed at $f_0$; the vertical dashed line marks the design frequency. This frozen-impedance sweep is a numerical sensitivity test and does not constitute a unit-cell bandwidth prediction.}
\label{fig:validation}
\end{figure*}

Figure~\ref{fig:validation}(a) shows convergence toward the 384-unknown result. Figure~\ref{fig:validation}(b) is a frozen-impedance sensitivity test, not a unit-cell bandwidth prediction. At $0.99f_0$ and $1.01f_0$, the worst protected-sector ratios are $-6.75$ and $-8.13$~dB, while directivity remains 12.61 and 12.56~dB. At $\pm2\%$, the low-scattering ratios degrade to approximately $-3.5$~dB. The design is therefore narrowband; a broadband claim would require dispersive state models and multi-frequency constraints.

To test state error, independent zero-mean Gaussian perturbations with 5\% standard deviation are applied to the resistance and reactance of every physical cell. Two hundred trials are solved on the 256-unknown mesh. A Park--Miller uniform generator followed by the Box--Muller transform, initialized with seed 271828, makes the perturbation sequence independent of Mathematica-version random-number conventions \cite{Mathematica9}. Table~\ref{tab:MC} gives empirical percentiles. The 95th-percentile worst-sector value remains $-7.81$~dB, telemetry directivity is tightly concentrated near 13.07~dB, and the sampled beam direction remains at $1^{\circ}$. These results do not replace a tolerance-aware unit-cell model, but they verify that the reported solution is not destroyed by small independent impedance errors.

\begin{table}[t]
\caption{Percentiles From 200 Five-Percent Impedance-Error Trials}
\label{tab:MC}
\centering
\footnotesize
\begin{tabular}{@{}lrrr@{}}
\toprule
Metric & 5th & 50th & 95th\\
\midrule
Worst sector (dB) & $-12.23$ & $-10.91$ & $-7.81$\\
Mean sector (dB) & $-14.24$ & $-12.78$ & $-11.28$\\
$\eta_{\mathcal M}$ & 0.634 & 0.646 & 0.655\\
SLL (dB) & $-14.01$ & $-13.48$ & $-12.95$\\
$D_{\mathrm{2D}}$ (dB) & 12.99 & 13.07 & 13.15\\
Pointing (deg) & 1.0 & 1.0 & 1.0\\
\bottomrule
\end{tabular}
\end{table}

\subsection{Independent Fourier--Galerkin Cross-Validation}

Mesh refinement within one MoM implementation tests discretization convergence but retains the pulse basis, midpoint testing, and logarithmic self-panel approximation. A second formulation was therefore implemented independently for the frozen eight-state library and the two frozen codebooks. It imports no MoM matrices, currents, fields, or output data. Only the geometry, impedances, codebooks, excitation definitions, and observation sectors are common. As detailed in Appendix~\ref{app:FG}, the circular single-layer operator is diagonalized in Fourier modes, while exact Fourier coefficients of the piecewise-constant cell admittance represent the discontinuous boundary profile.

\begin{figure*}[!tbp]
\centering
\includegraphics[width=1.00\textwidth]{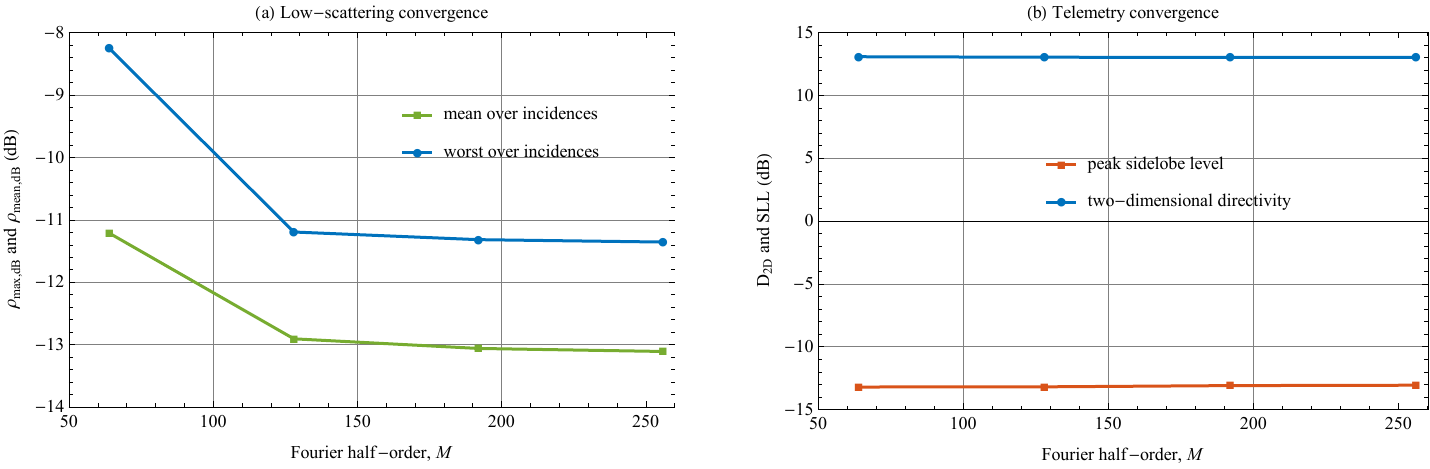}
\caption{Independent Fourier--Galerkin order convergence for the frozen shared-state design, where $M$ is the Fourier half-order and $2M+1$ modes are retained. (a) Worst and mean protected-sector power ratios. (b) Telemetry two-dimensional directivity and peak sidelobe level. The eight impedances and both 64-cell codebooks are held fixed, and no MoM matrices, currents, fields, or output data are used in this calculation. Stabilization with increasing $M$ provides an independent cross-formulation check of the MoM results; no impedance state or codebook is reoptimized.}
\label{fig:FGconvergence}
\end{figure*}

Figure~\ref{fig:FGconvergence} shows convergence for Fourier half-orders $M=64$, 128, 192, and 256, corresponding to 129 through 513 complex modes. Between the final two orders, the worst- and mean-sector metrics change by 0.038 and 0.048~dB, respectively; telemetry directivity and peak sidelobe level change by 0.0026 and 0.031~dB. The efficiency change is $3.60\times10^{-4}$, the sampled beam direction is unchanged, and all final projected relative residuals are below $5.6\times10^{-15}$.

Table~\ref{tab:crossValidation} compares the final Fourier--Galerkin values with both refined MoM results. The independent worst-sector value is within 0.22~dB of the 256-unknown MoM value and within 0.03~dB of the 384-unknown value. The larger differences in main-sector efficiency and peak sidelobe level remain within the declared audit tolerances and are consistent with the different representations of cell discontinuities. The absolute tolerances were fixed before the final comparison and were not used to retune the impedance states or codebooks. Every comparison passes. This agreement provides an independent cross-formulation check of the effective impedance-boundary solution. It does not supply unit-cell, bias-network, or platform-level validation.

\begin{table*}[t]
\caption{Independent Fourier--Galerkin Cross-Validation of the Frozen Shared-State Design}
\label{tab:crossValidation}
\centering
\footnotesize
\begin{tabular}{@{}lrrrrr@{}}
\toprule
Metric & MoM 256 & MoM 384 & FG, $M=256$ & FG$-$MoM 256 & Audit tolerance\\
\midrule
Worst protected-sector ratio (dB) & $-11.575$ & $-11.385$ & $-11.358$ & $+0.217$ & 1.00\\
Mean protected-sector ratio (dB) & $-13.089$ & $-12.879$ & $-13.111$ & $-0.023$ & 1.00\\
Main-sector efficiency & 0.6491 & 0.6435 & 0.6258 & $-0.0233$ & 0.040\\
$D_{\mathrm{2D}}$ (dB) & 13.098 & 13.082 & 13.018 & $-0.081$ & 0.25\\
Peak SLL (dB) & $-13.601$ & $-13.472$ & $-13.078$ & $+0.524$ & 0.75\\
Beam angle (deg) & 1.0 & 1.5 & 2.0 & $+1.0$ & 1.0\\
\bottomrule
\end{tabular}
\end{table*}

\section{Discussion and Scope}

The result supports four conclusions. First, the central contribution is the shared-state, dual-mode inverse problem: low scattering under several illuminations and directive telemetry can be made compatible with one finite passive impedance alphabet when the two modes are allowed different spatial codebooks. This coupling is absent from a collection of two separately optimized metasurfaces. Second, inverse-source null-space freedom provides an effective mechanism for solving that compatibility problem by moving among data-equivalent currents before common-state quantization. The near-machine-precision invariance in Fig.~\ref{fig:inverse} distinguishes this step from ordinary far-field optimization, but null-space exploitation by itself is not the novelty claimed here. Third, the small separate-library advantage in Table~\ref{tab:performance} indicates that sharing the state alphabet need not impose a large performance penalty. Fourth, the independent Fourier--Galerkin agreement in Table~\ref{tab:crossValidation}, together with MoM mesh refinement, shows that the principal conclusions are not artifacts of one basis or one self-panel approximation.

The limitations define the permissible interpretation. The study is two-dimensional, scalar, single-polarization, and single-frequency. The impedance boundary is local and isotropic; it does not model spatial dispersion, substrate thickness, finite conductivity beyond $R_0$, bias lines, switching transients, nonlinear loads, structural curvature variations, or temperature. No claim is made about plasma blackout, broadband stealth, or a flight-qualified skin. A cell design would need to map bias states to the dispersive impedances in \eqref{eq:finalStates}, and a three-dimensional vector integral equation would need to include polarization, bias-network interactions, and the actual platform. Thus, the results establish feasibility at the effective impedance-boundary level; they do not establish unit-cell, bias-network, or platform-level realizability. Conversely, the inverse formulation and proofs are not tied to the circle: replacing $\bm S$ and $\bm C$ by operators for an arbitrary triangulated surface leaves the shared-state, null-space, quantization, and perturbation steps unchanged.

The computational scaling is governed by the forward factorization and null-space construction. For large three-dimensional meshes, matrix-free products, randomized singular subspaces, adjoint derivatives \cite{LalauKeraly2013Adjoint}, and domain decomposition would replace dense algebra. The present dense models are intentionally small enough that both the MoM study and the independent Fourier--Galerkin calculation remain reproducible in Mathematica 9 \cite{Mathematica9} without proprietary full-wave software.

\section{Conclusion}

A dual-mode conformal metasurface has been cast as a shared-state inverse problem in which distinct low-scattering and telemetry codebooks must use one passive impedance alphabet. Inverse-source null-space freedom supplies the data-preserving mechanism used to reconcile the required currents with that common hardware description; it is a tool within the solution rather than the novelty by itself. The derivations establish the current-to-impedance Jacobian, optimal common-state update, monotonic inverse mismatch reduction, efficient rank-one forward update, and a stability bound. A mesh-refined cylindrical study and an independent Fourier--Galerkin formulation support the same conclusion: sectoral low scattering and directive telemetry are simultaneously feasible within the prescribed eight-state impedance-boundary model. The results establish feasibility at the effective impedance-boundary level; they do not establish unit-cell, bias-network, or platform-level realizability.

\section*{Acknowledgment}
M.R.K. gratefully acknowledges partial support from Prince Mohammad Bin Fahd University.
During preparation of the manuscript, the authors used ChatGPT (OpenAI) for drafting
assistance, language editing, and code review. The authors reviewed, edited, and approved all suggested changes and take full responsibility for the accuracy and originality of the content.

\section*{Data Availability}
The numerical data underlying the figures and the associated numerical study, together with the Wolfram Mathematica \cite{Mathematica9} source used to generate them, are available on GitHub \cite{Khodja2026RISCode}.

\appendices
\section{Analytic PEC-Cylinder Check}

For normal-incidence TM$_z$ illumination, the incident field expansion is
\begin{equation}
 e^{\jj kr\cos\phi}=\sum_{n=-\infty}^{\infty}\jj^nJ_n(kr)e^{\jj n\phi}.
\end{equation}
Enforcing zero total tangential electric field at $r=a$ gives
\begin{equation}
 E_z^{\mathrm{sca}}(r,\phi)=-\sum_{n=-\infty}^{\infty}
 \jj^n\frac{J_n(ka)}{H_n^{(1)}(ka)}H_n^{(1)}(kr)e^{\jj n\phi}.
 \label{eq:analyticPEC}
\end{equation}
The self-test truncates \eqref{eq:analyticPEC} at $\lvert n\rvert=80$ and compares it with \eqref{eq:singlelayer} evaluated from the numerical PEC current. The comparison tests the time convention, the sign in \eqref{eq:singlelayer}, the singular self term, and the linear solve simultaneously.

\section{Existence of the Finite-Dimensional Regularized Subproblem}

For completeness, restrict each null coefficient to a closed bounded set, use the passive compact interval \eqref{eq:passiveSet}, and let each codebook take values in the finite set $\{1,\ldots,K\}^N$. Let \(\epsilon_j\) be the prescribed positive current-floor threshold. After imposing the current floor $\lvert j_n\rvert\geq\epsilon_j>0$, the retrieval map \eqref{eq:retrieve} and the regularized objective are continuous. The product of the compact continuous-variable sets and the finite codebook sets is compact. The Weierstrass theorem therefore guarantees that the regularized finite-dimensional inverse stage attains a minimum. This existence result does not imply uniqueness; nonuniqueness is precisely the design freedom being exploited.

\section{Independent Fourier--Galerkin Formulation}
\label{app:FG}

The independent validation starts from the same continuous boundary equation as the MoM model but uses a global Fourier basis and Galerkin testing. For a $2\pi$-periodic boundary quantity $g(\phi)$, define
\begin{equation}
 g_n=\frac{1}{2\pi}\int_0^{2\pi}g(\phi)e^{-\jj n\phi}\,\dd\phi,
 \qquad
 g(\phi)=\sum_{n=-\infty}^{\infty}g_ne^{\jj n\phi}.
 \label{eq:FGseries}
\end{equation}
On the circle $r=a$, the single-layer operator in \eqref{eq:singlelayer} is diagonal in this basis. Indeed, the exterior limiting form of the cylindrical-wave addition theorem and Fourier orthogonality give
\begin{align*}
 \mathcal S[e^{\jj n\phi'}]
 &=\frac{k\eta a}{4}\sum_{q=-\infty}^{\infty}
 H_q^{(1)}(ka)J_q(ka)e^{\jj q\phi}\\
 &\quad\times\int_0^{2\pi}e^{-\jj q\phi'}
 e^{\jj n\phi'}\,\dd\phi'\\
 &=s_ne^{\jj n\phi},
\end{align*}
where the modal eigenvalues are
\begin{equation}
 s_n=\frac{\pi k\eta a}{2}J_n(ka)H_n^{(1)}(ka).
 \label{eq:FGeigenvalue}
\end{equation}
A direct finite expansion of $ZJ$ converges slowly at the 64 cell boundaries. The validation therefore uses the inverse-admittance factorization
\begin{equation}
 u=ZJ,\qquad Y=Z^{-1},\qquad J=Yu,
 \label{eq:FGfactorization}
\end{equation}
which transforms the boundary equation into $u+\mathcal S(Yu)=b$. Here, $y_p$ denotes the $p$th Fourier coefficient of the admittance
$Y(\phi)$, defined according to \eqref{eq:FGseries}. Retaining modes $-M\leq n,m\leq M$ gives
\begin{align}
 \sum_{m=-M}^{M}\left(\delta_{nm}+s_n y_{n-m}\right)u_m&=b_n,
 \label{eq:FGsystem}\\
 j_n&=\sum_{m=-M}^{M}y_{n-m}u_m.
 \label{eq:FGcurrent}
\end{align}
Every prescribed impedance is nonzero, so \eqref{eq:FGfactorization} is algebraically equivalent to the original continuous equation rather than an altered boundary model.

Let physical cell $\ell\in\{0,\ldots,63\}$ occupy $[\ell\Delta,(\ell+1)\Delta)$, where $\Delta=2\pi/64$. For a piecewise-constant cell quantity $v_\ell$, including $Y_\ell=1/Z_\ell$, its exact coefficient is
\begin{align}
 v_p&=\frac{1}{2\pi}\sum_{\ell=0}^{63}v_\ell
 \int_{\ell\Delta}^{(\ell+1)\Delta}e^{-\jj p\phi}\,\dd\phi\nonumber\\
 &=\frac{1}{64}\operatorname{sinc}\!\left(\frac{p\pi}{64}\right)
 \sum_{\ell=0}^{63}v_\ell e^{-\jj p(\ell+1/2)\Delta}.
 \label{eq:FGcellcoeff}
\end{align}
Here, $\operatorname{sinc}(x)=\sin(x)/x$ with its continuous value at $x=0$.
Thus, no midpoint approximation is used to represent the discontinuous impedance or admittance profile. The plane-wave and internal-line-source excitation coefficients are, respectively,
\begin{align}
 b_n^{\mathrm{pw}}&=\jj^nJ_n(ka)e^{-\jj n\alpha},
 \label{eq:FGplane}\\
 b_n^{\mathrm{line}}&=-\frac{k\eta}{4}H_n^{(1)}(ka)
 J_n(kr_f)e^{-\jj n\phi_f}.
 \label{eq:FGline}
\end{align}
After solving \eqref{eq:FGsystem}, the scattered far-field angular factor is evaluated from
\begin{equation}
 f^{\mathrm{sca}}(\theta)=-\frac{k\eta}{4}(2\pi a)
 \sum_{n=-M}^{M}(-\jj)^nJ_n(ka)j_ne^{\jj n\theta}.
 \label{eq:FGfarfield}
\end{equation}
For telemetry, the direct term
\begin{equation}
 f^{\mathrm{dir}}(\theta)=-\frac{k\eta}{4}
 e^{-\jj kr_f\cos(\theta-\phi_f)}
 \label{eq:FGdirect}
\end{equation}
is added before the metrics are formed.

The implementation evaluates sensitive Bessel--Hankel products at 50-digit precision before forming machine-precision dense systems. It verifies the exact cell coefficients against an independent 8192-point quadrature, the admittance multiplication matrix against pointwise multiplication and Fourier projection, a constant-impedance solution against its closed-form modal solution, and the modal PEC field against a 4096-point boundary quadrature. It also checks the projected relative residual
\begin{equation}
 r_M=\frac{\lVert\bm A_M\bm u_M-\bm b_M\rVert_2}
 {\lVert\bm b_M\rVert_2}
 \label{eq:FGresidual}
\end{equation}
for the frozen low-scattering and telemetry systems. Here, \(\bm A_M,\bm u_M,\bm b_M\) are the truncated Fourier–Galerkin system matrix, modal solution vector, and excitation vector, respectively. Production calculations use $M\in\{64,128,192,256\}$; the final order contains 513 complex modes. The convergence results, residuals, and comparison with the MoM values are reported in Fig.~\ref{fig:FGconvergence} and Table~\ref{tab:crossValidation}.

\IfFileExists{IEEEtran.bst}{\bibliographystyle{IEEEtran}}{\bibliographystyle{ieeetr}}
\bibliography{TAP_DualMode_Metasurfaces}

\end{document}